\documentclass[twocolumn,showpacs,preprintnumbers,amsmath,amssymb]{revtex4}
\usepackage{graphicx}
\input{epsf}

\begin{document}

\title{Surface photocurrent in electron gas   over liquid He subject to quantizing
magnetic field
}

\author{L.I. Magarill$^{+,*}$, M.V. Entin$^{+,*}$~\thanks{e-mail: entin@isp.nsc.ru}}
\address{$^{+}$Institute of Semiconductor Physics, Siberian
Branch of the Russian Academy of Sciences, Novosibirsk, 630090, Russia
\\ $^{*}$ Novosibirsk State University, Novosibirsk, 630090, Russia}
\begin{abstract}
The photogalvanic effect is studied in electron gas over
the liquid He surface with the presence of quantizing magnetic field.
The gas is affected by the weak alternating microwave electric field tilted towards the surface normal. Both linear
and circular photogalvanic effects are studied.  The current occurs via indirect
phototransition with the participation of ripplons emission or
absorption. The photogalvanic tensor has
strong resonances at the microwave frequency $\omega$ approaching to the frequencies of transitions between  size-quantized subbands.
 The resonances are symmetric or antisymmetric, depending on a tensor component.   Other resonances appear at  $\omega\approx n
\omega_c$, where  $n$ being integer and $\omega_c$ is the
cyclotron frequency.
It is found that the latter resonances split to two peaks connected with
 emission or absorption of ripplons.   The calculated photogalvanic coefficients are in accord with the experimental
observed values.\end{abstract}

 \maketitle

\subsection*{Introduction}
The stationery surface photocurrent (in other words, surface photogalvanic effect (SPGE)) appears along the border of isotropic homogeneous bounded medium under the action of tilted alternative electric field \cite{we4}, \cite{alper}. This effect was also studied in size-quantized systems \cite{we5}-\cite{we_jetpl2}. Recently, SPGE attracted attention \cite{chep} as a tentative source of  the microwave-induced photoresponse   oscillations in 2D electron gas over the liquid He surface (EGOHeS). The microwave-induced resistance oscillations (MIRO) were the subject of numerous publications (see, e.g., review \cite{dmitr}). In particular, this effect was studied theoretically \cite{Mon}-\cite{Mon3} in relation to EGOHeS (the theory of EGOHeS, see e.g., \cite{shik_mon}).

SPGE can be considered as another source of the observed photoresponse oscillations manifesting itself without stationary in-plane electric field. The theory of SPGE in EGOHeS was developed in \cite{we3} for the case of no magnetic field.

The present paper is a continuation of   \cite{we3} with accounting for a strong  magnetic field ${\bf B}$ directed perpendicular to the He surface $(x,y)$.   The forced progressive in-plane electron motion  in a quantizing magnetic field is a result of the  transitions between size-quantized subbands with synchronous directed in-plane transitions between the Landau states. The mathematical reflection of this idea is the second order optical transition probability with the participation of scattering, in particular, ripplon-induced scattering. The translational motion results from the interference of transition amplitudes caused by out- and in-plane components of alternating electric field. The scattering leads to a change of the in-plane electron momentum with a shift of the orbit center.

 We will mainly  follow the conditions of the experiment \cite{chep} on MIRO. We  consider the electron gas of low density ($\sim 10^6$ cm${}^{-2}$) over the He3 or He4 surface. At such low density the electron gas is non-degenerate. The photon energy is chosen close to   the  distance between the ground and the first excited size-quantized electron states $\Delta$. The magnetic field is assumed to be weak enough so that the cyclotron quantum is some times less than $\Delta$.

The mentioned resonance  works as a magnification factor for  transitions via the intermediate state. The mechanism of SPGE can be illustrated as follows. The population of subbands changes in- or contra-phase  with the normal component of the alternating field if the frequency exceeds or it is less than $\Delta/\hbar$ and has $\pi/2$ shift if $\hbar\omega=\Delta$. This fact, together with the phase shift of the  in-plane field component,  determines the current direction.

\subsection*{Problem formulation}

The phenomenology of SPGE in a magnetic field ${\bf B}={\bf  b}B\equiv B(0,0,b_z)$
is determined by the relation for the current density
\begin{eqnarray}\label{phenomen}\nonumber
   {\bf j}=\alpha_1\mbox{Re}\Big[({\bf E}-{\bf n}({\bf nE}))({\bf n
   E^*})\Big]-\frac{i\alpha_2}{2}[{\bf n},[{\bf E},{\bf E^*}]]+ \\\alpha_3[({\bf E}-{\bf n}({\bf nE})),{\bf b}]({\bf n
   E^*})+\frac{i\alpha_4}{2}[[{\bf n},[{\bf E},{\bf E^*}]],{\bf  b}],
\end{eqnarray}
where ${\bf n}$ is the outer normal to the quantum well, ${\bf
E}(t)=\mbox{Re}({\bf E}e^{-i\omega t})$ is the uniform microwave
electric field (${\bf E}=(E_x,E_y,E_z)$ is its complex amplitude). Real parameters $\alpha_i$ are the functions of magnetic field value $B$;
$\alpha_1$ and $\alpha_3$ correspond to linear and $\alpha_2$ and $\alpha_4$ - to circular photogalvanic effects, respectively. ''Drift'' components $\alpha_1$ and $\alpha_2$ exist in the case of zero magnetic field, while the ''Hall'' components $\alpha_3$ and $\alpha_4$ originate from the magnetic field action, change their signs with the magnetic field and vanish if ${\bf B}=0$.

The current components $\propto\alpha_2,~\alpha_4$ can be treated similar to translational motion of a rotating wheel. Electromagnetic field spin flow $ic[{\bf EE}^*]$ ($c$ is  speed of light) transfers its angular momentum to electrons as a moment of force. The friction converts this moment   to the translational electron motion.

We will base on the same model of 2D EGOHeS as in \cite{we3}.
Electrons are attracted to He via the dielectric image force and the normal static electric field which composes Coulomb-like states $\chi_l(z)$ with energies $\epsilon_l$.   The cyclotron frequency is supposed to be much lower than the Bohr energy.  The interaction of electrons with surface waves (ripplons) and the homogeneous alternating electric field  leads to the stationary surface current with density  ${\bf j}.$

 Eq.(\ref{phenomen}) for current contains additive contributions $\propto E_z E_x$ and $\propto E_z E_y$. The system under consideration is axially-symmetric. Hence, to determine all components of the photogalvanic tensor, one can find  only $x$ component of the current.

To calculate the photocurrent,  we will use the approach first suggested by Titeika \cite{titeica}. According to this approach, if unpertubed electron states are localized, the current can be expressed via the transition probability between  these states.
For the $x$-component of  the current density, one can write
\begin{eqnarray}
j_x=\frac{2e}{S}\sum_{\beta,\beta'}(X_\beta-X_{\beta'})W_{\beta\rightarrow\beta'}f(\varepsilon_\beta)(1-f(\varepsilon_{\beta'})),
\end{eqnarray}
where $W_{\beta\rightarrow\beta'}$ is the
 probability of transitions  (caused by perturbation) between the electron states with quantum numbers $\beta$  and $\beta'$, $ \varepsilon_\beta$ and $X_\beta$ are the energy and the center of localization, correspondingly, $f( \varepsilon_\beta)$ is the Fermi function,  $S$ is the  system area, $e$ is the electron charge.

 Let us choose   the vector potential of  magnetic field  in the form of ${\bf A}=(0,Bx,0)$ when  the electron states are localized in the $x$-direction. Electron states  are described by  a set of quantum numbers  $\beta=(l,n,k_y)$,  $l$ is the number of size quantized level, $n$ is the Landau number, $k_y$ is the $y$-component of electron momentum: $$|\beta\rangle=\frac{1}{\sqrt{S a}}e^{ik_yy}\phi_n(\frac{x-X_\beta}{a})\chi_l(z),$$  where $\phi_n(\xi)$ are dimensionless oscillator functions,   $X_\beta=-b_za^2k_y$ is the localization center  (the cyclotron orbit center), $a=\sqrt{c/|e|B}$ is the magnetic length (we set $\hbar=1$). These states have energies $ \varepsilon_\beta=\varepsilon_n + \epsilon_l$, ~ where  $\varepsilon_n=\omega_c(n+1/2)$ is the n-th Landau level ($n=0,1,...$),  $\omega_c=|e|B/mc$ is the cyclotron frequency,  $ \epsilon_l$ is the $l$-th size quantization level ($l=1,2,...$. If image attraction to liquid He prevails $ \epsilon_l=-1/(2ma_B^2 l^2)$, \begin{eqnarray}\label{chi}&&\chi_1(z)=2z\exp{(-z/a_B)}a_B^{-3/2}, \nonumber\\&&\chi_2(z)=z(2-z/a_B)\exp{(-z/2a_B)}(2a_B)^{-3/2},\end{eqnarray} where  $a_B=\kappa/me^2$ is the effective Bohr radius, $\kappa=4\kappa_1(\kappa_1+\kappa_2)/(\kappa_2-\kappa_1), \kappa_{1,2}$ are dielectric constants of gaseous  and liquid helium.
Note, that the functions (\ref{chi}) can be used  also in the presence of the normal static field, if to consider them as variational functions with fitting parameter $a_B$.

In our case  transition probability $W_{\beta\rightarrow\beta'}$ is determined by the  interaction with a microwave field and ripplons with the Hamiltonian  ${\cal H}_{int}(t) = {\cal U}_{er} + {\cal F}(t)$. The Hamiltonian of electron-ripplon interaction ${\cal U}_{er}$ is
\begin{equation}\label{Uer}
    ({\cal U})_{er} = S^{-1/2} \sum_{\bf q}J_{\bf q}(b_{-{\bf q}}^+ + {\bf q}_{\bf q})V_q(z),
    \end{equation}
where

\begin{equation}\label{Vq}
V_q(z)=\frac{1}{ma_B^3}\sqrt{\frac{q}{2 \rho}\omega_q}\bar{V}_q(z); ~~ \bar{V}_q(z) = \frac{a_B^2}{z^2}(1-qzK_1(qz)),
\end{equation}
$J_{\bf q}=e^{i{\bf qr}},~~{\bf r}=(x,y), ~~b_{\bf q}^+,~  b_{\bf q}$ are the operators of creation and destruction of ripplon with wave vector ${\bf q}$ and frequency $\omega_q=q^{3/2}\sqrt{\sigma_0/\rho}, \rho $ is is the liquid helium density, $\sigma_0$ is the helium surface tension coefficient.
Interaction of electron with microwave field is given by
\begin{eqnarray}
   {\cal F}(t)&=&\frac{ie}{2\omega} {\bf Ev}e^{-i\omega t} + h.c. \equiv \frac{1}{2}Ue^{-i\omega t}  + h.c.; \\
  U &=&\frac{ie}{\omega} {\bf Ev}=\frac{ie}{\omega} ({\bf E_\|v_\|}+E_zv_z),
\end{eqnarray}
where ${\bf v}$ is the operator of electron velocity.
 In the first order on the interaction Hamiltonian, the contributions of $ ({\cal U})_{er}$ and ${\cal F}(t)$ to the transition probability are additive and do not produce the photocurrent. Hence, the transition amplitude should be searched in the second (mixed in $ ({\cal U})_{er}$ and ${\cal F}(t)$) order. In this order we have
\begin{eqnarray}\label{Wph}\nonumber
W_{\beta\rightarrow \beta'}= \frac{\pi}{2S}\sum_{{\bf q},\pm}(N_{q}+\frac{1}{2}\pm\frac{1}{2})\times \nonumber \\ \Bigg[\delta(\varepsilon_{\beta,\beta'}-\omega \mp\omega_q)|({\cal B}_{\bf q})_{\beta\beta'}|^2 + \nonumber \\ \delta(\varepsilon_{\beta,\beta'}+\omega \mp\omega_q)|(\tilde{{\cal B}}_{\bf q})_{\beta\beta'}|^2\Bigg],
\end{eqnarray}
where
\begin{eqnarray} \label{B}
({\cal B}_{\bf q})_{\beta\beta'}=\sum_{\beta_1}\Bigg(\frac{(I_{\bf q})_{\beta,\beta_1}U_{\beta_1,\beta'}}{\eta-i\omega
+i\varepsilon_{\beta_1,\beta'}}+ \nonumber \\  \frac{U_{\beta\beta_1}(I_{\bf q})_{\beta_1,\beta'}}{\eta +
i\omega+i\varepsilon_{\beta_1,\beta}}\Bigg);~~~~~~~~~~~~~~(\eta = +0).
\end{eqnarray}
Here $I_{\bf q}=J_{\bf q}V_q(z), ~~ \varepsilon_{\beta,\beta'}$ means $\varepsilon_\beta -\varepsilon_\beta'$, $N_q$ is the ripplon equilibrium distribution function.
Quantity $(\tilde{{\cal B}}_{\bf q})_{\beta\beta'}$  is determined by Eq.(\ref{B}) with change $\omega\rightarrow -\omega$  and $U  \rightarrow U^+ $.
For matrix elements $(I_{\bf q})_{\beta,\beta'}$ one can write the following expressions:
\begin{eqnarray} \label{I}
(I_{\bf q})_{\beta,\beta'} = (J_{\bf q})_{n,k_y;n',k_y'}(V_q)_{l,l'},
\end{eqnarray}
 where
 \begin{eqnarray} \label{J}
 &&(J_{\bf q})_{n,k_y;n',k_y'} = \nonumber\\ &&\nonumber\delta_{X,X'-a^2q_y}i^{|n-n'|}e^{i(q_xX+b_zu_q\sin{(2\varphi)}/2)} e^{ib_z(n'-n)\varphi}J_{nn'}(u_q);\\
  &&J_{nn'}(u_q)=\sqrt{\frac{\min{(n,n')!}}{\max{(n,n')!}}}u_q^{|n-n'|/2}e^{-u_q/2}L_{\min{(n,n')}}^{|n-n'|},
 \end{eqnarray}
 $L_n^m(u)$ is the generalized Laguerre polynomial, $u_q=q^2a^2/2$, $\varphi$ is the polar angle of vector ${\bf q}~~ (q_x=q\cos{\varphi}, q_y=q\sin{\varphi}); ~~  (V_q)_{l,l'} = \int_0^\infty dz V_q(z)\chi_l(z)\chi_{l'}(z)$. We will consider the PGE at resonance conditions, when the microwave frequency is close to the distance between size-quantization subbands  $l=1$ and $l=2 $.
 The expressions for the necessary quantities $(V_q)_{1,1},~ (V_q)_{2,2}$ and $(V_q)_{1,2}$ can be found, for example, in \cite{we3}.

The matrix elements of operator $U$ are
\begin{eqnarray} \label{UU}
 &&U_{\beta,\beta'} =\frac{ie}{\omega}\delta_{X,X'}\Big[\delta_{l,l'}\frac{a\omega_c}{\sqrt{2}}\Big((iE_x+b_zE_y)
 \sqrt{n}\delta_{n,n'+1}+\nonumber\\ &&(-iE_x+b_zE_y)\sqrt{n'}\delta_{n',n+1}\Big)+\delta_{n,n'}E_z(v_z)_{l,l'}\Big]
 \end{eqnarray}

Using Eqs.(\ref{I}), (\ref{J})  and (\ref{UU}) we get for  PGE current:
\begin{eqnarray} \label{jx1}
 &&j_x=-\frac{2e}{2\pi a^2}\frac{e^2a^2b_z}{\omega^2}\frac{2\pi a \omega_c}{2\sqrt{2}S}\times\nonumber\\&&\sum_{{\bf q},\pm}\sum_{n,l;n',l'}f(\varepsilon_{n,l})(1-f(\varepsilon_{n',l'}))J_{n,n'} \big(N_q+\frac{1}{2}\pm \frac{1}{2}\big)q_y V_{l,l'}\nonumber \\
 &&\times 
 \Bigg\{\delta(\varepsilon_{n,n'}+\epsilon_{l,l'}-\omega\mp\omega_q) \times \nonumber \\
 &&\mbox{Re}\Big[E_z^*\sum_{l_1}\Big(\frac{V_{l,l_1}(v_z)_{l_1,l'}}{\eta-i\omega+\epsilon_{l_1,l'}}+
 \frac{(v_z)_{l,l_1}V_{l_1,l'}}{\eta+i\omega+\epsilon_{l_1,l}}\Big)^*\times \nonumber \\
  &&i\ \mbox{sig}(n'-n)\Big(e^{ib_z\varphi}E_{-}\frac{\sqrt{n'+1}J_{n,n'+1}-\sqrt{n}J_{n-1,n'}}{\omega_c-\omega}- \nonumber \\
   &&e^{-ib_z\varphi}E_{+}\frac{\sqrt{n'}J_{n,n'-1}-\sqrt{n+1}J_{n+1,n'}}{\omega_c-\omega}\Big)\Big] +\nonumber \\ &&(\omega\rightarrow -\omega, {\bf E}\rightarrow {\bf E}^*)\Bigg\},
  \end{eqnarray}
 where $E_{\pm}=E_x\pm ib_zE_y, \ \varepsilon_{n,n'}=\omega_c(n-n'), \ \epsilon_{l,l'}=\epsilon_{l}-\epsilon_{l'} $, $\mbox{sig}(x)=\mbox{sgn}(x+0)$.

  Quantities $J_{n,n'}$ justify the relations:
\begin{eqnarray} \label{J_rel}
(\sqrt{n'+1}J_{n,n'+1} - \sqrt{n} J_{n-1,n'})\mbox{sig}(n'-n) = \sqrt{u_q}J_{n,n'} ~~~\nonumber \\
 (\sqrt{n'}J_{n,n'-1} - \sqrt{n+1}J_{n+1,n'})\mbox{sig}(n'-n) =\sqrt{u_q}J_{n,n'}.~~
 \end{eqnarray}
  Using Eq.(\ref{J_rel}) one can rewrite Eq.(\ref{jx1}) in the form
  \begin{eqnarray} \label{jx2}
 &&j_x=\frac{e^3a \omega_c}{2\sqrt{2}\pi \omega^2}\sum_{\pm}\sum_{n,l;n',l'}f(\varepsilon_{n,l})(1-f(\varepsilon_{n',l'}))\times \nonumber \\ &&\int\limits_0^\infty dqq^2J_{n,n'}^2 \sqrt{u_q}\big(N_q+\frac{1}{2}\pm \frac{1}{2}\big) V_{l,l'}\times\nonumber\\&&
 \Bigg\{\delta(\varepsilon_{n,n'}+\epsilon_{l,l'}-\omega\mp\omega_q) \times \nonumber \\
&& \mbox{Re}\Big[\sum_{l_1}\Big(\frac{V_{l,l_1}(v_z)_{l_1,l'}}{\eta-i\omega+\epsilon_{l_1,l'}}+
 \frac{(v_z)_{l,l_1}V_{l_1,l'}}{\eta+i\omega+\epsilon_{l_1,l}}\Big)^*E_z^*\nonumber\\&&\times\frac{\omega_cE_x-ib_z\omega E_y}{\omega_c^2-\omega^2}\Big]  + (\omega\rightarrow -\omega, {\bf E}\rightarrow {\bf E}^*)\Bigg\}.
   \end{eqnarray}

Because of  low electron concentration,  function $f(\varepsilon)=e^{(\mu -\varepsilon)/T}$  is the Boltzmann distribution function ($\mu$ being  the chemical potential), $f \ll 1$.  We  will consider the resonance case when microwave frequency is close to the distance between subbands with $l=1$ and $l=2$. Assuming that $\Delta \gg T$ and  leaving only resonance terms one can reduce Eq.(\ref{jx2})  to:
\begin{eqnarray} \label{jx3}Å
&&j_x=\frac{n_ee^3z_{12}\Delta a(1-e^{-\omega_c/T})}{2\sqrt{2}m \omega^2(\omega_c^2-\omega)}\sum_{\pm,n,n'}e^{-n\omega_c/T}\int\limits_0^\infty dqq^2J_{n,n'}^2  \nonumber \\ &&\times\sqrt{u_q}\big(N_q+\frac{1}{2}\pm \frac{1}{2}\big) V_{1,1} \mbox{Im}\Bigg\{E_z^*(\omega_cE_x-ib_z\omega E_y) \times  \nonumber \\
&& \Big[\frac{V_{1,2}}{\eta-i\delta}\delta(\omega_c(n-n')-\omega\mp\omega_q) +\nonumber \\ &&
\frac{V_{1,2}}{\eta+i\delta}\delta(\omega_c(n-n')+\omega\mp\omega_q) + \nonumber \\ &&\Big(\frac{V_{1,1}}{\eta-i\delta}+\frac{V_{2,2}}{\eta+i\delta}\Big)\delta(\omega_c(n-n')-\delta\mp\omega_q)\Big] \Bigg\}.
   \end{eqnarray}
   Here $\Delta=\epsilon_2-\epsilon_1,~ \delta=\Delta-\omega$ is the resonance detuning, $n_e$ is the electron density.

   At the fulfilment of condition $\omega_c \gg T$ Eq.(\ref{jx3}) can be simplified:
   \begin{eqnarray} \label{jx4}Å
   &&j_x=\frac{CF}{\delta^2+\eta^2}\Big[\mbox{Re}(E_xE_z^*)\omega_c\delta+\mbox{Im}(E_xE_z^*)\omega_c\eta-\nonumber \\ &&\mbox{Re}(E_yE_z^*)b_z\omega\eta+ \mbox{Im}(E_yE_z^*)b_z\omega\delta\Big],  \end{eqnarray}
    where
   \begin{eqnarray}
F=\sum_{n} \Big[\big(N_q+1\big)\theta(\omega - n\omega_c) + N_q\theta(n\omega_c-\omega)\Big]\times \nonumber\\  \Big(\frac{\omega_q}{\omega}\Big)^{4/3}\frac{1}{n!}u_q^ne^{-u_q}\bar{V}_{1,1}(qa_B)\bar{V}_{1,2}(qa_B)\Bigg|_{\omega_q=|\omega - n\omega_c|},~~
   \label{F}\\
  C=\frac{n_ee^3z_{12}\Delta a^2 \rho^{2/3}}{12 m^3 a_B^6 \omega^{2/3}(\omega^2-\omega_c^2)\sigma_0^{5/3}},~~~~~~~~~~~~~~~~~~~~~~~~~\end{eqnarray}
    $u_q=\omega_q^{4/3}(\rho/\sigma_0)^{2/3}a^2/2,  ~~ qa_B=  \omega_q^{2/3}(\rho/\sigma_0)^{1/3}a_B.$

    For $\bar{V}_{ij}(y)$  we have\\
$ \bar{V}_{11}(y)=2y^2(y^2-4)^{-3/2}\Big[(y^2-4)^{1/2}-2\arccos(2/y)\Big],$
$\bar{V}_{12}(y)= (8y^2\sqrt{2}/9)(4y^2-9)^{-5/2}\Big[(4y^2-9)^{1/2}(9+8y^2)-36y^2\arccos(3/(2y ))\Big].$

    The comparison of Eq.(\ref{phenomen}) with Eq.(\ref{jx4}) gives for the photogalvanic coefficients $\alpha_i=CF a_i$:
    \begin{eqnarray}\label{pge_coef}
         a_1=\frac{\omega_c\delta}{\eta^2+\delta^2}, ~a_2=\frac{\omega_c\eta}{\eta^2+\delta^2}, \nonumber \\  ~ a_3=\frac{\omega\eta}{\eta^2+\delta^2}, ~ a_4=\frac{\omega\delta}{\eta^2+\delta^2}.
    \end{eqnarray}
        Eq.(\ref{F}) has been obtained neglecting the Landau level widths. We have remained the  widening $\eta$ of the intersubband distances in the prefactors $a_i$ only to get the finite result.  To include the Landau level widths, one should blur the delta-functions under the integral in Eq.(\ref{F}) as $\delta(x)\to \eta_1/(x^2+\eta_1^2)/\pi$. In principle, widths $\eta$ and $\eta_1$ may be different, but here we will suppose  $\eta_1=\eta$.

Thus, the dependence of the photogalvanic coefficients on the magnetic field is mainly determined by the function $F$. Fig. 1 shows this function in the case of zero electron level widths. For numerical calculations we used the parameters of He3 and electron gas close to the conditions of experiment \cite{chep}: $\kappa_1 = 1, \kappa_2 = 1.057$,   $\rho=0.082$g/cm${}^{-3}$, $\sigma_0=0.1553$erg/cm${}^{-2}$ and $n_e=1.4\cdot 10^6$cm${}^{-2}$. The parameter $a_B=8.7\cdot 10^{-7}$cm is chosen to fit the intersubband distance  $\Delta =0.38$ meV corresponding to the experiment \cite{chep}. The dependence of $F(B)$ contains narrow twin peaks in the vicinity of cyclotron resonance harmonics $\omega=n\omega_c$.  At high temperature,  the left and right peaks have the same amplitudes. With the drop of temperature, the right  peak   is suppressed as compared to the left one. The way the widening of electron levels affect the  shape of resonances is demonstrated in Fig.2 - Fig.4.

\begin{figure}[h]\label{fig1}\centerline{\epsfxsize=10cm\epsfbox{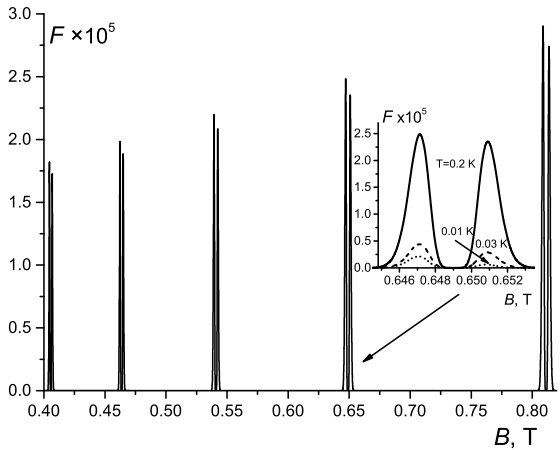}}
\caption{Function $F$ {\it versus} magnetic field at $T=0.2$K. The widening of the Landau levels is neglected.  The double peaks correspond to $\omega=n\omega_c$ for $n=8,7,6,5,4$ (from left to right). Insert:  the fine structure of e double peak  $n=5$ at different temperatures. }

\end{figure}

\begin{figure}[h]\label{fig2}\centerline{\epsfxsize=10cm\epsfbox{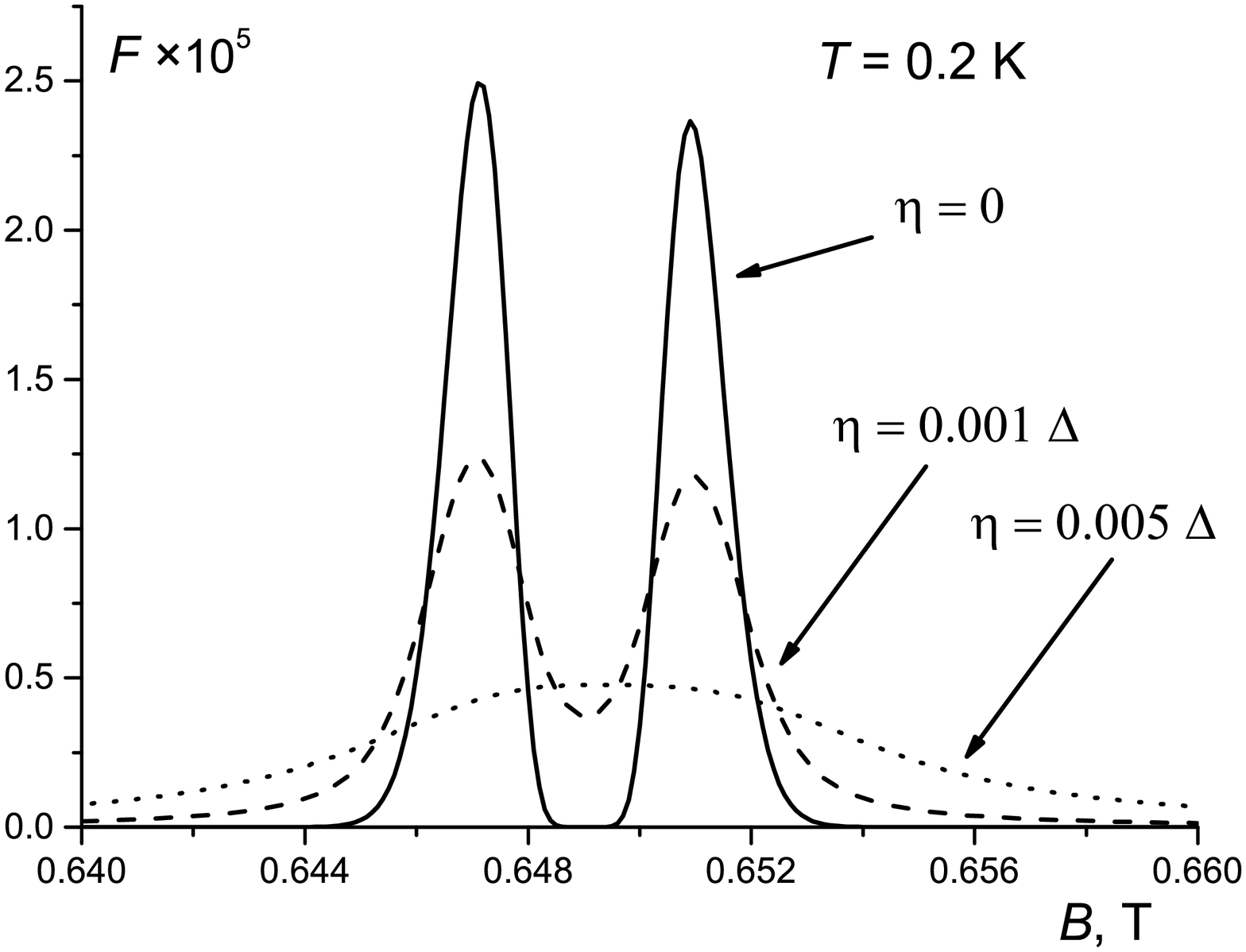}}
 \caption{Fig.2. The evolution of $F$ {\it versus} magnetic field with the  change of the Landau level width in the region of peak  $\omega/\omega_c=5$. The widening smears the twin structure. }
\end{figure}
\begin{figure}[h]\label{fig3}\centerline{\epsfxsize=10cm\epsfbox{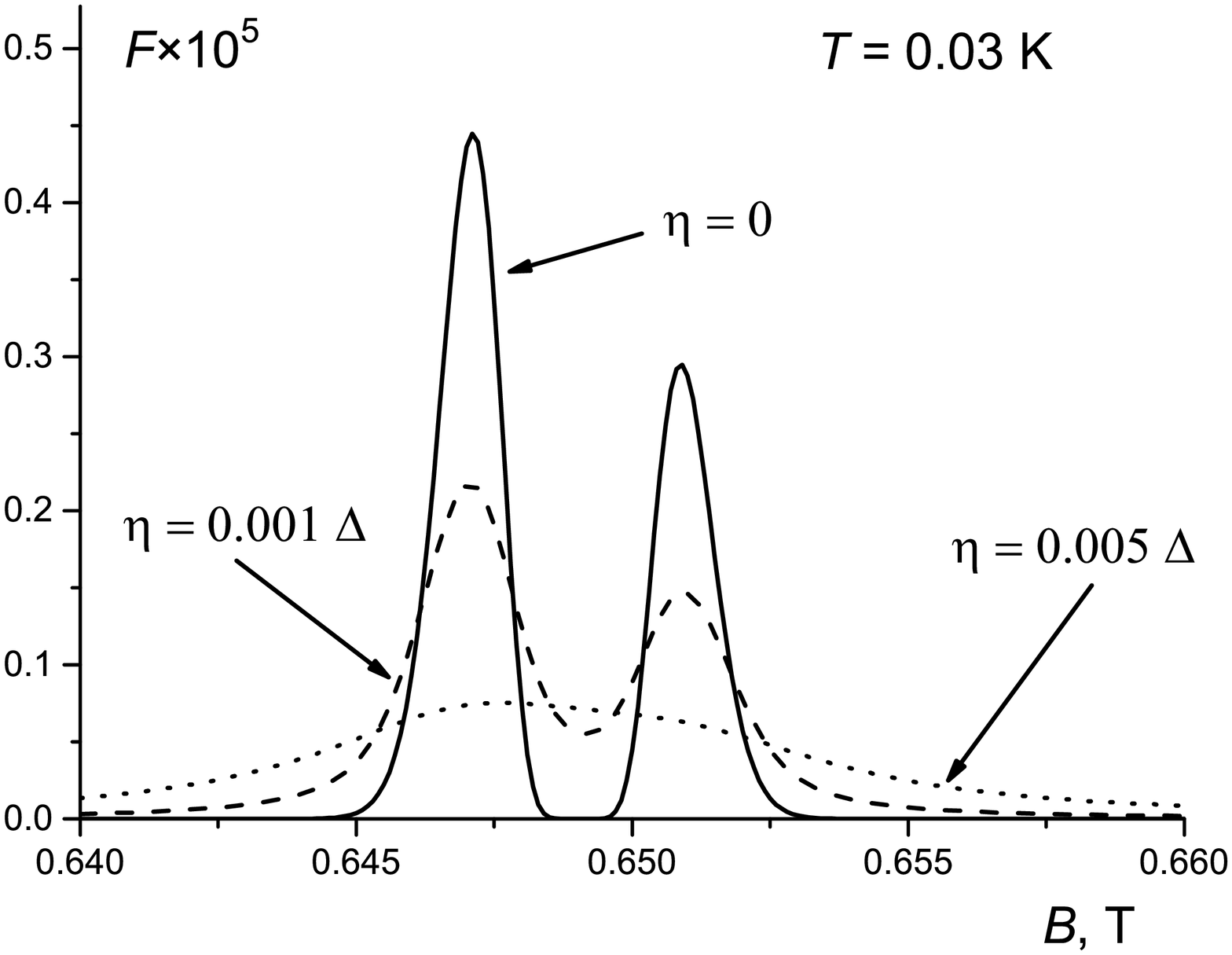}}
 \caption{Fig.3. The same as in Fig.2  for $T=0.03$K. }
\end{figure}
\begin{figure}[h]\label{fig4}\centerline{\epsfxsize=10cm\epsfbox{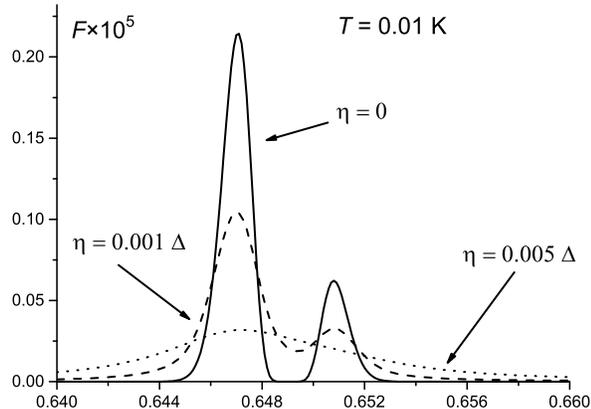}}
 \caption{Fig.4. The same as in Fig.2  for $T=0.01$K.}
\end{figure}

\subsection*{Analysis of results}
The photocurrent dependence on the parameters has a huge ''zoo'' of resonances. First, this is the resonance $\hbar\omega\approx\Delta$ in coefficients $a_i$  (see Eq.(\ref{pge_coef})). This resonance originates from the participation of $E_z$ component in transitions and it is not specific for the case with magnetic field.  According to our previous papers, this resonance appears due the intermediate state. It determines the symmetric or antisymmetric behavior of PGE coefficients in the vicinity of  resonance frequency, depending on the kind of
  electromagnetic field polarization (linear or circular). In addition to SPGE coefficients existing also at $B=0$ \cite{we3},  the magnetic field leads to the Hall photocurrent component appearance.

Another resonance is the cyclotron resonance when the frequency of external field coincides with the cyclotron frequency. In the domain of parameters which we concentrate here this resonance is far from our focus (and the experimental situation in \cite{chep}).

One more kind of resonances is the cyclotron resonance harmonics $\omega\approx n\omega_c$ (corresponding to magnetic fields $B_n=m\Delta c/en$) on which we focus our attention (see Fig.1). These resonances originate from the participation of the scattering in the transition processes induced by alternating electric field.  Scattering violates the linear character of cyclotron motion resulting in cyclotron harmonics. Such resonances are typical for MIRO.

According to Eq. (\ref{pge_coef}), the maxima (as functions of detuning $\delta)$ of the Hall components of the photogalvanic tensor $\alpha_3, ~\alpha_4$ exceed corresponding  drift components $\alpha_1, ~\alpha_2$:  $\max_\delta(\alpha_1): \max_\delta(\alpha_2):\max_\delta(\alpha_3):\max_\delta(\alpha_4)=0.5:1:n:0.5n$, where $\omega=n\omega_c$. This is typical situation for transport in a strong magnetic field  where the drift along the drawing force is weaker than in the Hall direction. The maximal values of symmetric resonances are 2 times larger than antisymmetric ones.

In fact, these resonances are twinned (see insert to Fig.1). Note, that in quasielastic approximation the resonances should be solitary. For example, this is the case when impurity scattering is take into account only. In the considered situation of ripplon scattering the resonance splitting originates from the energy of emitted or absorbed ripplon $\omega_q=|n\omega_c-\omega|$. At $B\to B_n$, $\omega_q\to 0$, hence the process  probability vanishes. In particular, if the temperature is relatively high, $T>\omega_q$,  $F(B)\propto|B-B_n|^{4n/3+3}\ln(1/|B-B_n|)$. This explains  deep dips between the twin peaks at $n\omega_c=\omega$.

On the other hand, the peaks splitting can be estimated as $\sqrt{\sigma_0/\rho}(2n+9/2)^{3/4}n^{-7/4}(m\Delta)^{3/4}2mc/e$. This estimate is consistent with insert to Fig 1. The left and right peaks in pairs correspond to ripplons emission/absorption, accordingly. Their height ratio is $(N_q+1)/N_q$. At a $T\gg \omega_q$ the peaks heights equalize and behave $\propto T$; when $T\to 0$ both peaks go down, but the left one tends to constant, while the right one is suppressed $\propto \exp{(-\omega_q/T)}$.

In  the aforesaid specific conditions corresponding to \cite{chep} the parameter $C$ has the value $C=-88$ pA$\cdot$cm/V${}^2$ at $n=5$. If to suppose the level width $\eta=10^{-4}\Delta$ and $T=0.2$K this yields maximum of photogalvanic coefficient $\max_\delta(\alpha_3)=2.5\max_\delta(\alpha_1)=5\max_\delta(\alpha_2)=2.5\max_\delta(\alpha_4)=2.2$pA$\cdot$cm/V${}^2$ achieved at magnetic field $B=0.647$T. These values are commensurable with the order of MIRO oscillations in \cite{chep}.

In conclusion, we have found the value of photocurrent along the charged liquid helium surface  affected by  tilted alternating electric field in the presence of vertical magnetic field. Different photogalvanic coefficients represent responses to the linear and circular polarization of drift (invariant with respect to the sign of magnetic field) and Hall (proportional to this sign) currents. The ripplon scattering  mechanism was taken into account. SPGE coefficients have (symmetric or antisymmetric) their resonant behavior when the field frequency approaches  the intersubband one. Besides, resonances  on the cyclotron harmonics are observed.   The current value  is consistent with that observed in the experiment.

\subsection*{Acknowledgements}
The authors are grateful to A. Chepelyanskii for fruitful discussion.  This research was supported  by RFBR grants No 13-0212148 and No 14-02-00593.

 \end{document}